# Hybrid Classical-Quantum Deep Learning Models for Autonomous Vehicle Traffic Image Classification Under Adversarial Attack

**Reek Majumder**
Glenn Department of Civil Engineering, Clemson University, Clemson, SC
Email: rmajumd@clemson.edu, ORCID: 0000-0002-4659-7506

**Sakib Mahmud Khan, PhD**
Assistant Research Professor, Glenn Department of Civil Engineering
Clemson University, Clemson, SC
Email: sakibk@clemson.edu

**Fahim Ahmed**
Department of Civil and Environmental Engineering, University of South Carolina
300 Main St, Columbia, SC 29208
Email: ahmedf@email.sc.edu

**Zadid Khan, PhD**
Glenn Department of Civil Engineering, Clemson University
Clemson, SC 29631, USA
Email: mdzadik@clemson.edu

**Frank Ngeni**
Department of Engineering, South Carolina State University
Orangeburg, SC 29117,USA
Email: fngeni@scsu.edu

**Gurcan Comert, PhD**
Comp. Sci., Phy., and Engineering Department, Benedict College
1600 Harden Street, Columbia, SC 29204
Email: Gurcan.Comert@Benedict.edu

**Judith Mwakalonge, PhD**
Department of Civil and Mechanical Engineering Technology and Nuclear Engineering
South Carolina State University
Email: jmwakalo@scsu.edu

**Dimitra Michalaka, PhD**
Department of Civil and Environmental Engineering, The Citadel
Charleston, SC 29409, SC
Email: Dimitra.Michalaka@citadel.edu

**Mashrur Chowdhury, PhD**
Eugene Douglas Mays Chair in Transportation
Glenn Department of Civil Engineering
Clemson University, Clemson, SC 29631, USA
Email: mac@clemson.edu

**ABSTRACT**

Image classification must work for autonomous vehicles (AV) operating on public roads, and actions performed based on image misclassification can have serious consequences. Traffic sign images can be misclassified by an adversarial attack on machine learning models used by AVs for traffic sign recognition. To make classification models resilient against adversarial attacks, we used a hybrid deep-learning model with both the quantum and classical layers. Our goal is to study the hybrid deep-learning architecture for classical-quantum transfer learning models to support the current era of intermediate-scale quantum technology. We have evaluated the impacts of various white box adversarial attacks on these hybrid models. The classical part of hybrid models includes a convolution network from the pre-trained Resnet18 model, which extracts informative features from a high dimensional LISA traffic sign image dataset. The output from the classical processor is processed further through the quantum layer, which is composed of various quantum gates and provides support to various quantum mechanical features like entanglement and superposition. We have tested multiple combinations of quantum circuits to provide better classification accuracy with decreasing training data and found better resiliency for our hybrid classical-quantum deep learning model during attacks compared to the classical-only machine learning models.

**INTRODUCTION**
In recent years, the transportation system has been revolutionized by computing and communication innovations and technology. The aim of developing these technologies is to make driving safer and easier by sensing the surrounding environment. Nowadays, vehicles are equipped with sensors like radar, lidar, and cameras, where data collected by these sensors are processed to build vehicle guidance and driver assistance systems. The application of machine learning (ML) techniques, such as object classification *(1)*, face recognition *(2)*, object detections *(3) (4)*, motion recognition *(5)*, and target tracking *(6)*, have become an integral part of the autonomous vehicle development process. For example, the continuous images gathered by the in-vehicle cameras are used to detect roadways, signs, and obstacles. Thus, the development of vision-based techniques such as image classification is crucial for automotive technology. Traffic signs provide valuable information to navigate on the roadway. For autonomous driving, perceiving the traffic signs and taking actions accordingly is mandatory. Various ML-based image classification techniques have been used to classify traffic sign images, specifically speed limit and stop sign. The typical image classification model involves extracting features from images and applying machine learning models where real-time traffic detection is paramount to improving traffic safety *(7)*. Researchers reported difficulties in using convolutional neural networks models in analyzing traffic sign images due to: i) processing complexity of images having traffic signs in a narrow area of the image (*8*) with other surrounding objects (pedestrian, parked vehicles) (*9*), and ii) computational time required for complex and sophisticated image processing. Often researchers use public databases (*10–12*) to develop the models; however, these databases often do not capture real-world situations like illumination variations, weather, image deformations, and sign color deteriorations.

Apart from the image classification model difficulties, the performance of the ML model can significantly deteriorate with adversarial attacks, where perturbations injected in the data can fool the deep learning models to predict incorrect outputs (*13–16*). Earlier studies have been conducted to develop robust mechanisms to solve these attacks (*15,17,18*). Recent development in classical-quantum machine learning (CQ-ML) models could provide a potential solution toward a resilient model for image classification under adversarial attack. It provides a solution to the limited computation capabilities of classical computers (*19*). Quantum computing is becoming more mainstream with time. Attackers may leverage quantum computing to breach the cyber-physical system, which can be expensive on classical computers to tackle. One of the benefits of quantum computing is that users can test and develop their quantum programs in their local computers using quantum simulators before using actual quantum computers through the cloud *(20, 21)*. Various companies like Google *(22)*, IBM *(20)*, and Microsoft Azure *(23)* develop and provide access to real quantum computers to test for quantum-enabled ML(Q-ML) models for resiliency against adversarial attacks. However, it is still in its development phase and unreliable in practical application. Hybrid classical-quantum models can help explore quantum computers' possibilities *(20, 22–24)*.

The research objectives are to: (i) design and develop hybrid classical-quantum machine learning models to make them resilient against adversarial attacks, and (ii) compare the performance of classical and hybrid classical-quantum machine learning models for images under different adversarial attacks. Five adversarial attacks models for traffic sign images are used to measure and compare the accuracy between the neural network model and quantum machine learning models for classification.

## LITERATURE REVIEW

Improvement in hardware performance has always supported innovation in the computing community. The introduction of low-cost GPUs from 2010 has triggered various researches in image recognition (*25*) and object detection (*26*) using deep learning models. Before GPUs, various handcrafted features such as scale-invariant features transform (SIFT) (*27*) and histogram of oriented gradients (HOG) are designed and combined to build a bag-of-features (BoF)(*28*) for conventional machine learning models and used for practical application of image classification. These features were based on algorithms and biased on dataset and developer. Deep learning approaches have helped to generalize and automate feature extraction for the image recognition task using a convolutional neural network as the building block for these models.

Deep learning models and architecture supported by GPUs have contributed towards general object recognition and image classification tasks required for autonomous vehicles. Various models like AlexNet (*29*), VGG (*30*), and Inception (*31*), and ResNet (*25*) for image classification, designed to improve the traditional convolutional neural networks (*32*) for feature extractor and image classification tasks. Due to the lack of a large dataset and massive training time and resources required by these models. Most of these models are extended using the Transfer Learning (*33*) methodology of ML where a trained model on the ImageNet (*34*) dataset with 1.2 million images for 1000 categories. The convolution layer of these models is frozen and acts as a feature extractor for our desired dataset. Finally, the fully connected layer of these models is replaced with custom layers of neural networks designed for classification and used to fine-tune the model for newer datasets.

Although these models provide higher accuracy for the classification task, they are still vulnerable to various adversarial attacks. These attacks are designed to deceive machine learning models into producing erroneous predictions. The types of adversarial attacks on machine learning models depend mainly on two attributes: attackers' goal and attackers' knowledge of the model. The attackers' intent can be to perform a targeted attack or a non-targeted attack. In a targeted attack scenario, the attacker intends the adversary to misinterpret in a certain way. The non-targeted attack occurs when the attacker does not care about the prediction if the final results are incorrect. In the context of attackers' knowledge, two types of attacks can be performed: White box attack and Black box attack. The Limited Memory Broyden-Fletcher-Goldfarb-Shanno (L-BFGS) algorithm is an early strategy to fool the neural network models for image recognition. The goal of L-BFGS is to find a perceptually minimal input perturbation to deceive the model. An alternative formulation to the L-BFGS algorithm is the Adversarial Manipulation of Deep Representations (AMDR). The optimization is performed in the intermediary layers rather than in the output layers as L-BFGS. This method optimizes the similarity between a perturbated source image and a target image with a different class label.

The Fast Gradient Sign Method (FGSM) (*35*) is a single-step gradient ascent attack strategy. The FGSM method uses a hyper-parameter to control the amplitude of disturbance applied to the original sample. A few variants of FGSM were introduced in the literature. For example, One-Step Target Class Method (OSTC) (*36*), Basic Iterative Method (BIM) (*37*), Projected Gradient Descent (PGD) (*38*), and Iterative Least-likely Class Method (ILLCM) (*39*). The OSTC maximizes the probability of a class for prediction, which is less likely than the original sample. The BIM method applies the FSGM method multiple times to generate adversarial examples. The ILLCM method is a variant of the BIM method for a targeted attack. In this method, the target class is determined from the true class using the original data to have the least likelihood of being chosen. All these attacks modify most, if not all, input features. A few attack strategies only consider sparse perturbation like the Jacobian-based Saliency Map (JSMA) (*41*) and One Pixel Attack (*42*). The saliency map finds the influence of each input feature on models' class prediction. Thus, JSMA perturbates a small set of influential features to cause misclassification. The one-pixel attack only changes the value of one pixel of an image to cause misclassification. An iterative process is used to choose that one pixel for the best attack effect. In a DeepFool attack, the classifier estimates the distance of decision boundaries around a data point. The attack allows the classifier to go beyond the decision boundaries to misclassify the prediction. The HOUDINI (*43*) deceives the gradient-based algorithms by generating adversarial samples specific t the task loss function.

In this research, our hybrid models include circuit-based quantum layers as a part of the neural network. The Quantum Layer of "Parameterized Quantum Circuits," also referred to as Variational circuits. These circuits are designed to train circuit-based quantum models like neural networks. They are comprised of three main phases, Data Embedding Phase (*44*), Variational circuit phase (*45*), and quantum measurement. Data Embedding embeds data from the previous layer into quantum states. 'Variational circuit' phase uses various parameterized gates and entanglement operations to improve our classification task. And finally, a quantum measurement is used to read the quantum states and use it as an input to the following classical or quantum layer.

## METHOD

### Hybrid and Classical Image Classification Models

Since it is relatively rare to have a dataset large enough to train a Convolutional Network from scratch, transfer learning (36) is one of the most widely used machine learning techniques for image classification and object detection tasks. The goal of transfer learning is to re-use trained deep convolutional neural network models like ResNet (*25*), VGG (*30*), AlexNet (*41*) on large datasets like ImageNet (34), which contains over 1.2 million images of 1000 categories on newer problem statements.

For our experiment, we implemented transfer learning using a pre-trained ResNet18 model trained on ImageNet Dataset. Using the weights of this model's convolution network helps us use the model as a feature extractor. Later, we have fine-tuned the model by replacing the fully connected layer of the Resnet18 with a quantum layer and classical layer-enabled PyTorch (*39*) module. It is tested with the classical PyTorch module with a two-layer neural network. Our classical model's custom hidden layer uses a modified rectilinear unit (ReLU6) (*46*) as a nonlinear activation function (*47*), which helps neural networks to learn complex data patterns. We haven't used any explicit activation function for our hybrid model because the quantum measurement to read data from quantum computers provides us with that functionality. The fully connected layer for the Resnet model is replaced and trained with our new classical and hybrid model for evaluation.

For our analysis, we have used ResNet18 from TorchVision and generate a module for classical neural networks and classical-quantum neural networks using PyTorch. A Quantum circuit is developed using PennyLane (*48*), and quantum computation is simulated using the default Qubit device provided by PennyLane. To integrate the quantum PennyLane circuit with the classical PyTorch linear layer, the PennyLane TorchLayer is used, which is provided in version PennyLane v0.15.0.

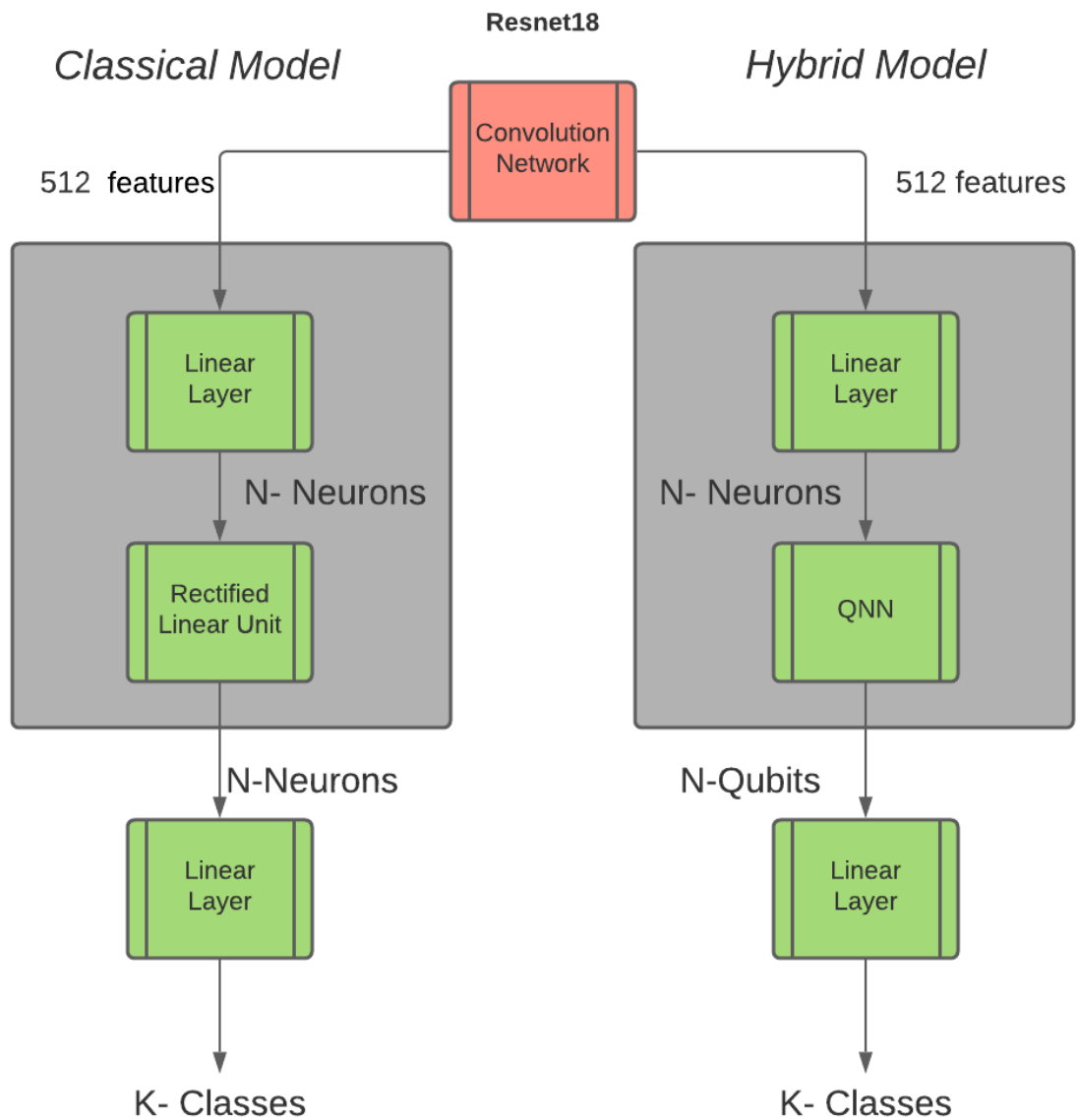

**FIGURE 1** Classical and Hybrid Model Architecture.

**Algorithmic steps to Quantum Circuits**

As shown in Figure 2, there are three steps in the quantum layer:

1) Data Embedding Layer: Embeds classical data into quantum states by combining one or more single qubit quantum gates like Hadamard gate, Rotational Y, Rotational X, Rotational Z, U1, U2, and U3 gates.

2) Variational Circuit Layer: Parameterized circuits are designed using two-qubit gates (Controlled NOT, Controlled Z, and Controlled RX) with single qubits parameterized gates (i.e., Rotational Y, Rotational X, Rotational Z, U1, U2, U3).
3) Measurement Layer:- Quantum measurement component is used to read the current state of the qubit. The results from the quantum layer are transferred to the next layer as classical data. For experimentation, we have used X, Y, and Z basis.

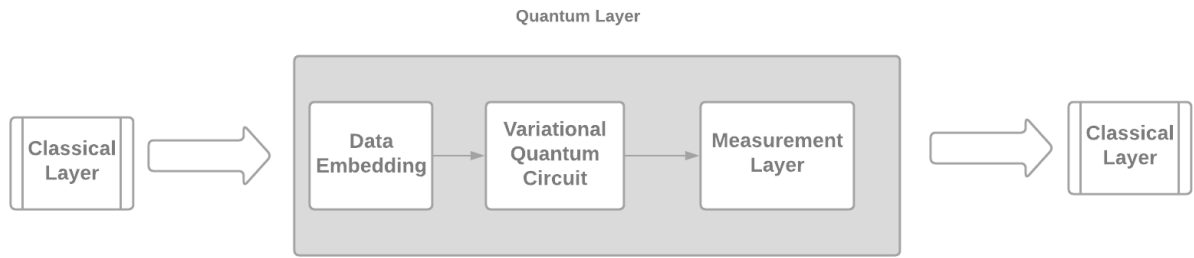

**FIGURE 2** Algorithmic steps for Quantum Circuit generation

**Image Dataset**

To study the performance of hybrid and classical models, we have used a subset of the extended LISA *(49)* traffic sign dataset, which contains around 7,855 annotations from 6,610 video frames classifying 47 different traffic signs. Image frames vary from $640 \times 480$ to $1024 \times 522$ pixels. The size of annotation boxes for traffic signs ranges between 6 x 6 to 167 x 168 pixels.

The number of samples for each type of traffic sign differs significantly, and we modified the traffic sign dataset into 18 traffic signs by focusing on the traffic sign and preprocessing images (crop and reducing the noise in its surroundings). The binary classification model used only stop signs and coded as one if stop sign, zero for all other signs. For multiclass, three types of roadway signs were considered: stop signs, speed limit signs, and other signs. For binary classification models, the LISA dataset had 231 samples with an 80/20 split for training and testing the models (number of training samples:182, number of testing samples: 49). For multiclass classification models, the LISA dataset had 279 samples with an 80/20 split for training and testing the models (number of training samples: 222, number of testing samples: 57). In addition, to test whether the hybrid Q-ML model performs better than the classical model with limited available data, binary classification models were trained with 40% of data and tested with 60% of data (i.e., 40/60 split).

**Attack Models**

Adversarial attacks on the classical ML model and hybrid ML model can help to study the model resiliency. Figure 3 shows a scenario where the image classifier model is $C$ and the victim sample is $(x, y)$ where output $y$ the target associated with input $x$. In the attack scenario, the attacker devises a fake input by using a perturbation to the original information such that it is perceptually similar to the actual image $x$. This fake sample misleads the classifier $C$ and provides us $y_p$ (y- predicted), which is not equal to the original class $y_o$ (y-original).

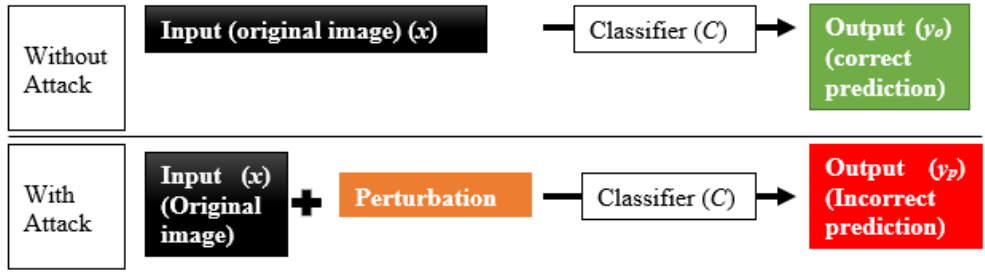

**FIGURE 3** Attack process to cause misclassification

In this study, five types of attack models used are: the gradient attack, Fast Gradient Sign Method (FGSM), Projected Gradient Descent (PGD) attack, Sparse L1 Descent Attack, and Simultaneous Perturbation Stochastic Approximation (SPSA). These are white-box attacks where attackers have access to model parameters except SPSA, a black box attack where the attacker has no access to the model parameters. These attacks differ on how the perturbation is applied to the original input to generated misclassified output. The gradient attack perturbs the input with the gradient of the model's loss function for the input. Instead of only using a gradient, the FSGM uses the gradient sign with a fixed magnitude to the input to implement the perturbation. The PGD obtains adversarial examples by using the fast gradient method iteratively, and the iteration starts with uniformly randomly chosen data inside the $L_2$ norm region near data. A sparse adversarial attack perturbs various parts of the input image randomly. In addition, the black box attack SPSA used in this study simultaneously perturbs input data in all dimensions. An example from the attack on the stop sign is in Figure 4.

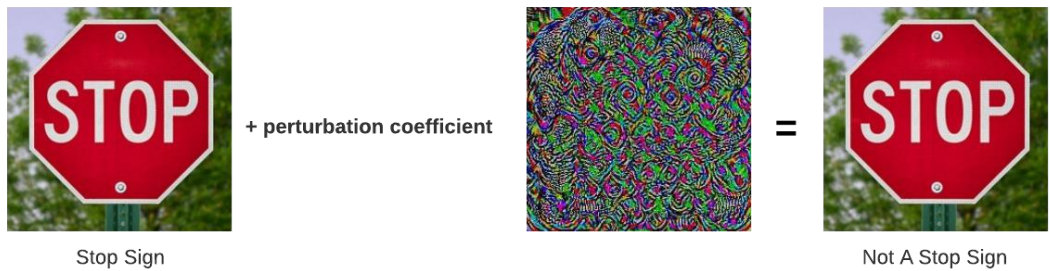

**FIGURE 4** Example of adversarial attack on stop signs

**MODEL DEVELOPMENT, ANALYSIS, AND DISCUSSION**

**Hybrid Model 1**

The quantum circuit for Hybrid model 1 uses four qubits for binary classification. These qubits are introduced with Rotational Y (Ry) gate as the data embedding layer that takes inputs from the previous layer of the neural network. The parameterized quantum circuit is designed using entanglement with a Controlled Z (CZ) gate with a combination of Rotational Y (Ry) and Rotational Z (Rz) gate, which is repeated for six iterations. Finally, we measure the results from the quantum layer in the Z-basis state, as shown in Figure 5.

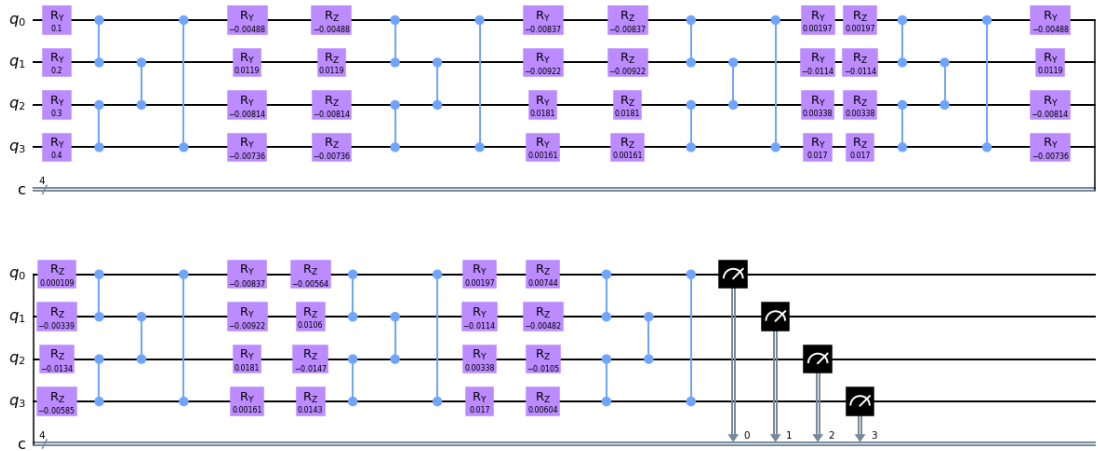

**FIGURE 5** Hybrid Model 1 for binary classification with randomly assigned initial weights

**Hybrid model 2**

The quantum circuit for Hybrid model 2 uses four qubits for binary classification. These qubits are introduced with the Hadamard gate to bring it into a superposition state. These states are treated with Rotational Y (Ry) gate as the data embedding layer that takes inputs from the previous layer of the neural network. The parameterized quantum circuit is designed using entanglement with a Controlled Z (CZ) gate with a combination of Rotational Y (Ry) and Rotational Z (Rz) gate, which is repeated for six iterations. We rotate the rotational X (Rx) gate with the input parameters from the previous layer and the Hadamard gate. Finally, measure the results from the quantum layer in the Z-basis state, as shown in Figure 6. We have extended this model for multiclass classification using six qubits as shown in Figure 7

We have used the cross-entropy loss function to maintain the symmetry among the hyperparameters for classical and quantum models, with adam optimizer and a learning rate of 0.004.

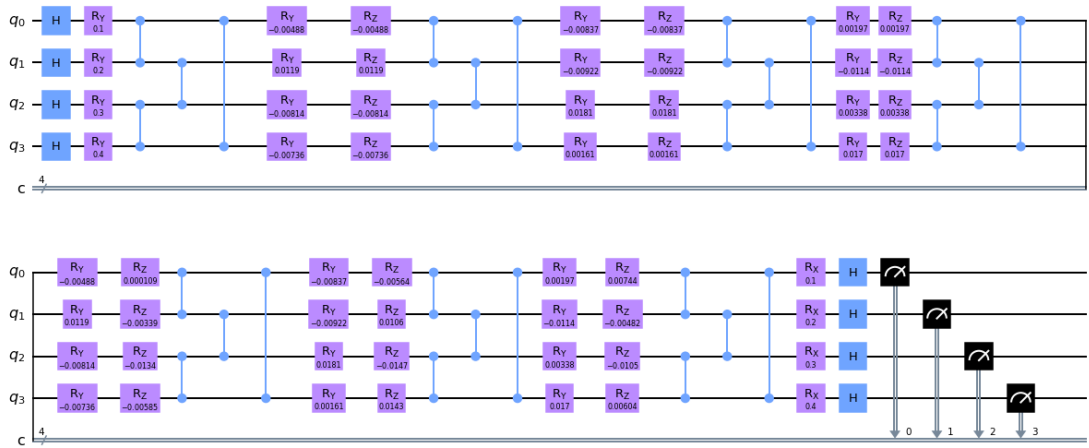

**FIGURE 6** Hybrid Model 2 for binary classification with randomly assigned initial weights

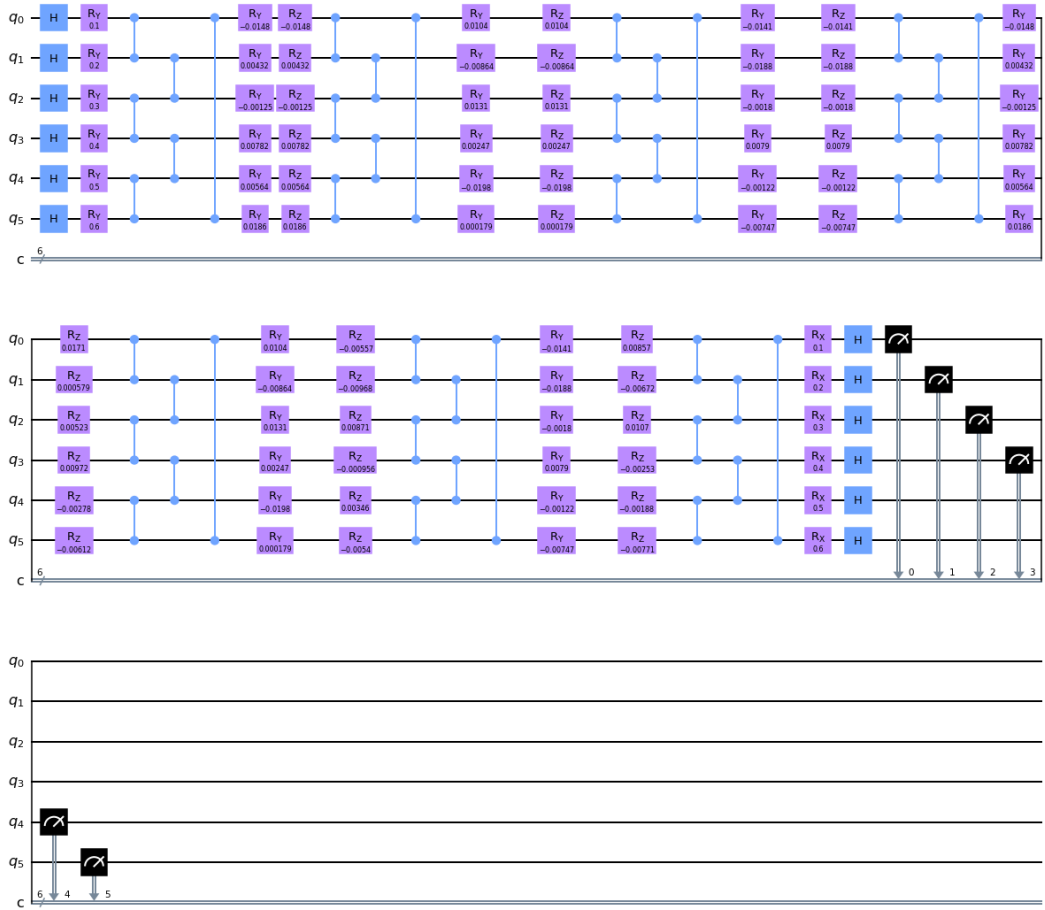

**FIGURE 7** Hybrid Model 2 for multiclass classification with randomly assigned initial weights

**Model Evaluation**

Tables 1 and 2 presents the performance of classical ML and hybrid Q-ML models in terms of classification accuracy of stop signs. A split of 80/20 (train/test) was used to develop these models. As shown in Table 1 for without attack scenario of stop sign image classification, the accuracy of the Hybrid 2 model was higher than the classical and Hybrid 1 model for both perturbation coefficients of 0.05 and 1. When the perturbation coefficient is 0.05, the Hybrid 2 models' accuracy was higher than classical and Hybrid 1 model for gradient, sparse L1 descent attack, SPSA, L2 PGD attack. The accuracy of the Hybrid 1 model was higher than the classical and Hybrid 2 model for both perturbation coefficients of 0.05 and 1 for gradient sign attack. In all attack cases for binary classification, the accuracy of Hybrid Q-ML models (Hybrid 1 and 2) was higher than the classical ML model. Table 2 shows the results for multiclass classification. As shown in Table 2, the accuracy of the Hybrid 2 model was higher than the classical and Hybrid 1 model for both perturbation coefficients of 0.05 and 1 in without attack scenario. The Hybrid 2 model has higher accuracy than the classical and Hybrid 1 model for all attack cases except for gradient sign attacks.

These results suggest that the hybrid Q-ML models perform better than classical models under adversarial attack scenarios for image classification. Specifically, Hybrid model 2 has been shown to provide better classification accuracy and resiliency during Gradient Attack, Sparse L1 Descent Attack, SPSA Attack, and L2 PGD attack.

**TABLE 1** Accuracy (%) of binary classification models (80/20 split)

| Attack Model | Perturbation Coefficient | Classical | Hybrid Model 1 | Hybrid Model 2 |
|---|---|---|---|---|
| Without attack | 0.05 | 91.30 | 93.47 | **100.00** |
|  | 1.0 | 91.30 | 93.47 | **100.00** |
| Gradient Attack | 0.05 | 89.13 | 89.13 | **91.30** |
|  | 1.0 | 89.13 | 89.13 | 89.13 |
| Gradient Sign Attack | 0.05 | 86.95 | **89.13** | 82.60 |
|  | 1.0 | 80.43 | **82.60** | 71.73 |
| Sparse L1 Descent Attack | 0.05 | 89.13 | 89.13 | **91.30** |
|  | 1.0 | 89.13 | 89.13 | 89.13 |
| SPSA Attack | 0.05 | 89.13 | 89.13 | **91.30** |
|  | 1.0 | 89.13 | 89.13 | 89.13 |
| L2 PGD Attack | 0.05 | 89.13 | 89.13 | **91.30** |
|  | 1.0 | 89.13 | 89.13 | 89.13 |

*Bold numbers indicate hybrid models with the highest accuracy in the related test event*

**TABLE 2** Accuracy (%) of multiclass classification models (80/20 split)

| Attack Model | Perturbation Coefficient | Classical | Hybrid Model 2 |
|---|---|---|---|
| Without attack | 0.05 | 98.18 | **100** |
|  | 1.0 | 98.18 | **100** |
| Gradient Attack | 0.05 | 92.72 | **96.36** |
|  | 1.0 | 90.90 | **94.54** |
| Gradient Sign Attack | 0.05 | 92.72 | **92.72** |
|  | 1.0 | **56.36** | 50.90 |
| Sparse L1 Descent Attack | 0.05 | 92.72 | **96.36** |
|  | 1.0 | 92.72 | **96.36** |
| SPSA Attack | 0.05 | 92.72 | **96.36** |
|  | 1.0 | 92.72 | **96.36** |
| L2 PGD Attack | 0.05 | 92.72 | **96.36** |
|  | 1.0 | 92.72 | **96.36** |

*Bold numbers indicate hybrid models with the highest accuracy in the related test event*

**Model Evaluation with Limited Available Data**

To understand Hybrid Q-ML model's performance on limited available data, the models were trained with 40% of the data and tested with 60% of data for binary classification. The results are shown in Table 3. As shown in Table 3, the accuracy of the Hybrid 2 model was higher than the classical and Hybrid 1 model for both perturbation coefficients of 0.05 and 1 in without attack scenario. The Hybrid 1 model has higher accuracy than the classical and Hybrid 2 model for all attack scenarios. These results suggest that the hybrid Q-ML model performs better than the classical ML models under adversarial attack and when trained with limited data.

**TABLE 3 Accuracy of binary classification models (40/60 split)**

| Attack Model | Perturbation Coefficient | Classical | Hybrid Model 1 | Hybrid Model 2 |
|---|---|---|---|---|
| Without attack | 0.05 | 92.02 | 94.20 | **95.65** |
|  | 1.0 | 92.02 | 94.20 | **95.65** |
| Gradient Attack | 0.05 | 85.50 | **92.75** | 91.30 |
|  | 1.0 | 85.50 | **92.75** | 91.30 |
| Gradient Sign Attack | 0.05 | 86.23 | **92.02** | 90.57 |
|  | 1.0 | 70.28 | **76.81** | 73.18 |
| Sparse L1 Descent Attack | 0.05 | 85.50 | **92.75** | 91.30 |
|  | 1.0 | 85.50 | **92.75** | 91.30 |
| SPSA Attack | 0.05 | 85.50 | **92.75** | 91.30 |
|  | 1.0 | 85.50 | **92.75** | 91.30 |
| L2 PGD Attack | 0.05 | 85.50 | **92.75** | 91.30 |
|  | 1.0 | 85.50 | **92.75** | 91.30 |

*Bold numbers indicate hybrid models with the highest accuracy in the related test event*

## CONCLUSIONS

Machine Learning models used by autonomous vehicles for decision-making in a real-time scenario. However, these models are vulnerable to various kinds of adversarial attacks, adding perturbation to the input data. These attacks can compromise the safety of autonomous vehicles. Having resilient ML models, which are not trained with the adversarial dataset but still able to classify transportation infrastructure with malicious data, will make the system more reliable and expedite autonomous vehicle deployment.

Due to the known theoretical advantages of quantum computation and the current revolution of quantum hardware, quantum-enabled algorithms can enhance the performance of the classical machine learning models. We tested the performance of the classical-quantum transfer learning models. We found that these models can maintain model performance above 90% during adversarial attacks. Implementation of classical-quantum transfer learning model architecture can help design new machine learning models that can improve the model's performance and resiliency for autonomous vehicles in image classification

Further work needs to be done to evaluate the effects of environmental and physical system errors in quantum computing on the classification tasks. In the future, we will assess quantum algorithms that can support the classical-quantum model for multi-class classification and real-time object detection while providing robustness towards more adversarial attacks.
,

## ACKNOWLEDGMENTS

This study is supported by the Center for Connected Multimodal Mobility (C2M2) (a US Department of Transportation Tier 1 University Transportation Center) headquartered at Clemson University, Clemson, South Carolina, USA. Any opinions, findings, conclusions, and recommendations expressed in this material are those of the author(s). They do not necessarily reflect the views of C2M2, and the US Government assumes no liability for the contents or use thereof.

**AUTHOR CONTRIBUTIONS**

The authors confirm contribution to the paper as follows: study conception and design: MC, RM, SK, ZK, FA, FN; data collection: RM, ZK, SK, MC; analysis and interpretation of results: RM, ZK, SK, MC; draft manuscript preparation: SK, RM, FA, FN, ZK, MC, GC, DM, JM. All authors reviewed the results and approved the final version of the manuscript.

1646. Dahl GE, Sainath TN, Hinton GE. Improving deep neural networks for LVCSR using rectified linear units and dropout. In: 2013 IEEE International Conference on Acoustics, Speech and Signal Processing. 2013. p. 8609–13.
47. Sibi P, Jones SA, Siddarth P. ANALYSIS OF DIFFERENT ACTIVATION FUNCTIONS USING BACK PROPAGATION NEURAL NETWORKS. . Vol. 2005;47:5.
48. Bergholm V, Izaac J, Schuld M, Gogolin C, Alam MS, Ahmed S, et al. PennyLane: Automatic differentiation of hybrid quantum-classical computations. ArXiv181104968 Phys Physicsquant-Ph [Internet]. 2020 Feb 13 [cited 2021 Aug 1]; Available from: http://arxiv.org/abs/1811.04968
49. Møgelmose A, Liu D, Trivedi MM. Detection of U.S. Traffic Signs. IEEE Trans Intell Transp Syst. 2015 Dec;16(6):3116–25.